\documentclass[aps,prl,twocolumn,showpacs,floatfix]{revtex4}
\usepackage{graphicx,epsfig,amsmath,amssymb}

\begin{document}

\title{The 3D structure of the Lagrangian  acceleration in turbulent flows}
\author{Nicolas Mordant}
\altaffiliation{present address: Laboratoire de Physique Statistique, Ecole Normale Sup\'erieure, 24 rue Lhomond, 75005 Paris, France.}
\author{Alice M. Crawford}
\author{Eberhard Bodenschatz}
\affiliation{Laboratory of Atomic and Solid State Physics, Clark Hall, Cornell University, Ithaca NY}

\begin{abstract}
We report experimental results on the three dimensional Lagrangian acceleration in highly turbulent flows. Tracer particles are tracked optically using four silicon strip detectors from high energy physics that provide high temporal and spatial resolution. The components of the acceleration are shown to be statistically dependent. The probability density function (PDF) of the acceleration magnitude is comparable to a log-normal distribution. Assuming isotropy, a log-normal distribution of the magnitude can account for the observed dependency of the components. The time dynamics of the acceleration components is found to be typical of the dissipation scales whereas the magnitude evolves over longer times, possibly close to the integral time scale. 
\end{abstract}


\maketitle

The Lagrangian approach has been fruitful in advancing of the understanding of the anomalous statistical properties of turbulent flows~\cite{Falkovich01}. Lagrangian experimental studies lagged behind theory because both high spatial and high temporal resolution are required for Lagrangian measurements. It has been possible only recently to obtain well resolved measurements (e.g.~\cite{Ott00,Laporta01,Mordant01e,Voth02,Crawford}). These experiments raised a renewed theoretical interest in the Lagrangian approach. For example, anomalous scaling properties may be related to non-extensive statistical mechanics~\cite{Beck01,Mordant04b}. Recently, refined stochastic models have been proposed to reproduce the observed intermittency of turbulence~\cite{Sawford03,Reynolds03}. However, there remains a lack of experimental data to confront theory and models. At Cornell University, we have developed a high speed three dimensional (3D) imaging system to study the statistics and dynamics of tracer particles. Here, we report data on the 3D structure of acceleration, the correlations between the acceleration components, and the statistics of the acceleration magnitude. We show that the acceleration magnitude probability density function (PDF) can be described by a log-normal distribution.
Assuming isotropy, the shape of the acceleration component PDF can be derived. The dynamical difference between the components and the magnitude is discussed.


The flow and the detectors have been described in detail in a previous article~\cite{Voth02}. The flow was of the Von K\'arm\'an type, driven by two coaxial disks with blades and rim, 20~cm in diameter, 33~cm apart. The water was enclosed in a cylindrical tank with a diameter of 48.3~cm. The disks were rotated at the same angular velocity but in opposite sense. 
\begin{figure}
\centering
\includegraphics[width=7cm]{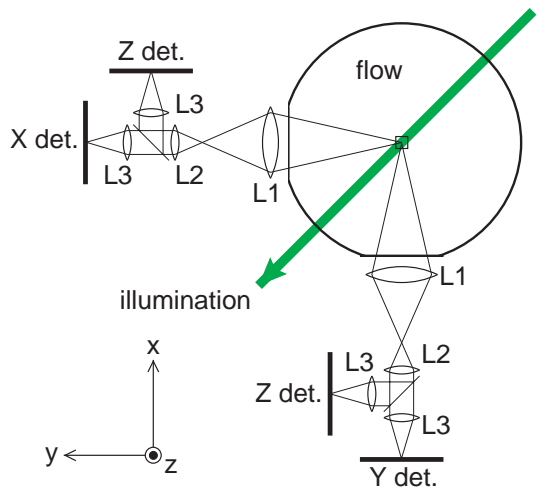}
\caption{Top view of the setup of the imaging system. A small measurement volume of size $4\times4\times2$~mm$^3$ at the center of the flow was illuminated by a 35 W pulsed laser. This volume was then imaged at a 45$^\circ$ scattering angle in the mid plane of the cylindrical tank, so as to record the three coordinates of the particle motion in the flow. The characteristics of the optics were identical to those in~\cite{Voth02}. The rotation axis of the disks was along the $z$ direction.}
\label{fig-setup}
\end{figure}

A schematic of the imaging system is displayed in Fig.~\ref{fig-setup}. A 35~W pulsed YAG laser illuminated the center of the flow. A $4.1\times4.1\times2.05$~mm$^3$ volume was imaged onto 4 silicon strip detectors. { This is a major improvement on the previous setup which had only two detectors and thus yielded only 1D or 2D tracks~\cite{Voth02})}. Each detector was made of 512 strips (pixels) and recorded one component of the position. The particles were polystyrene spheres, 25~$\mu$m in diameter, with density 1.06 times that of water. They were shown to behave as neutral tracer particles~\cite{Voth02}.

Since each detector recorded only one coordinate, the four recordings had to be matched to build a 3D track. The algorithm used to process the raw data has already been described in~\cite{Voth02}. For each pair of detectors ($(x,z)$ or $(y,z)$), the intensity scattered by the particles as they moved through the laser beam  was highly correlated. The intensity signal was used to match the $x$ and $z$ coordinates or the $y$ and $z$ coordinates in each pair. The last step consisted of matching the two recordings of the $z$ component to get the full 3D trajectory. To compute the acceleration from the tracks, the position signal was convolved with a Gaussian kernel that both differentiated and filtered the noise. { In this way, we were able to record, for the first time, large data sets of resolved 3D acceleration in highly turbulent flows}.

\section{Statistics of the acceleration vector}

\begin{figure}
\includegraphics[width=8.5cm]{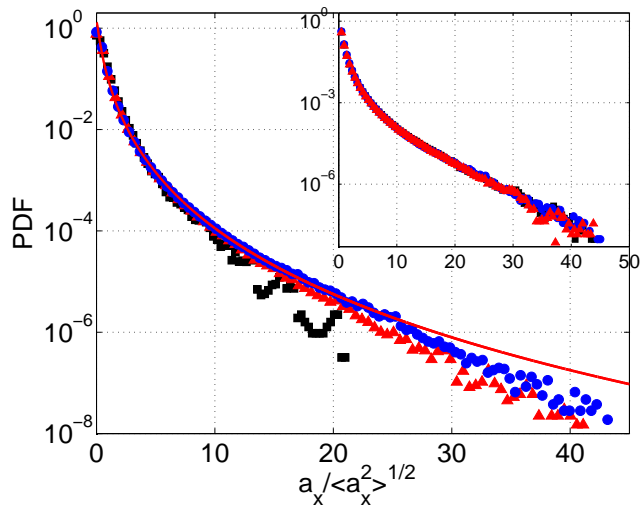}
\caption{PDF of one acceleration component at various Reynolds numbers: $R_\lambda=285$ ({\small $\blacksquare$}), 485 ({\large $\blacktriangle$}) and 690 ({\Large $\bullet$}). Only positive values of the acceleration are shown. Solid line: PDF of a component assuming the magnitude is log-normal with variance 1. Insert: PDFs of the 3 acceleration components at $Re=690$.}
\label{fig-pdfacc}
\end{figure}
The insert of Fig.~\ref{fig-pdfacc} shows the normalized PDFs  of the three acceleration components for $R_\lambda=690$. The three curves collapse showing that the shape of the PDF is the same for all directions. Nevertheless, the acceleration is not isotropic as the variances of the axial and transverse components are different. The ratio of axial to transverse acceleration variance decreases from 1.17 at $R_\lambda=285$ to 1.06 at $R_\lambda=690$~\cite{Voth02}. As can be seen from Fig.~\ref{fig-pdfacc}, the tails of the PDF are increasingly wide as the Reynolds numbers grows.
This confirms the results previously reported in~\cite{Voth02}. The tails are very wide, meaning  that the Lagrangian acceleration is a very intermittent quantity. The flatness was quite hard to estimate: its very high value required large data sets and the presence of noise made the estimation difficult~\cite{Crawford,Mordant04b}. { Here we took advantage of the two simultaneous recordings of the $z$ component, giving two independent realizations of the noise. In this way we got a new estimation of the value of the flatness, less sensitive to the noise.  
For the highest Reynolds number measured, the flatness was slightly over 100.}

\begin{figure}[!htb]
\includegraphics[width=8cm]{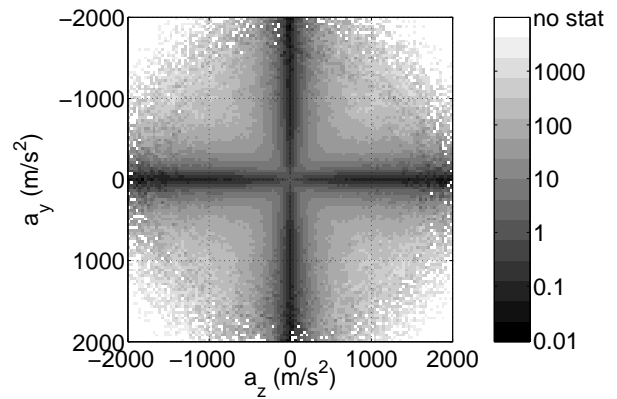}
\caption{Ratio $\frac{P(a_y,a_z)}{P(a_y)\,P(a_z)}$. The Reynolds number is $R_\lambda=690$.}
\label{fig-jPDF}
\end{figure}
Most stochastic models for Lagrangian dispersion in isotropic and homogeneous turbulence are one dimensional and model the three components of acceleration independently (e.g.~\cite{Sawford91}). In reality, the shape of the PDF of the components implies that the components are not independent since the only possible distribution for an independent and isotropic acceleration vector is Gaussian.
One way to check the dependence of the acceleration components is to study the ratio $\frac{P(a_1,a_2)}{P(a_1)\,P(a_2)}$, which is 1 for independent variables. This quantity is shown in Fig.~\ref{fig-jPDF}(b). The joint probability of simultaneously observing two components taking high absolute values is higher than the product of the probabilities by more than three orders of magnitude. On the contrary, it is much less probable to observe one component taking a small value and another a large absolute value. 
The simultaneous occurrence of large values of different acceleration components is in agreement with previous observations of Mordant {\it et al.} in the inertial range~\cite{Mordant02b}. 

\begin{figure}[!htb]
\includegraphics[width=8.5cm]{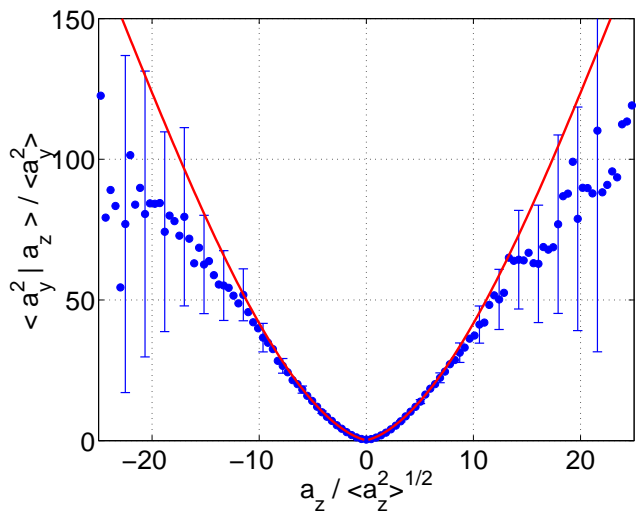}
\caption{Conditional variance $\langle a_y^2|a_z\rangle$ at $R_\lambda=690$ ({\Large $\bullet$}). Solid line: prediction from a log-normal distribution of the magnitude of variance 1 (see text).}
\label{fig-jPDFmoycond}
\end{figure}
Another way to check the dependence of the components is to compute the variance of one component conditioned on another. Figure~\ref{fig-jPDFmoycond} shows $\langle a_y^2|a_z\rangle$ at $R_\lambda=690$. The conditional variance increases strongly with $|a_z|$, taking values larger than 50 times the full acceleration component variance.


\begin{figure}
\includegraphics[width=8.5cm]{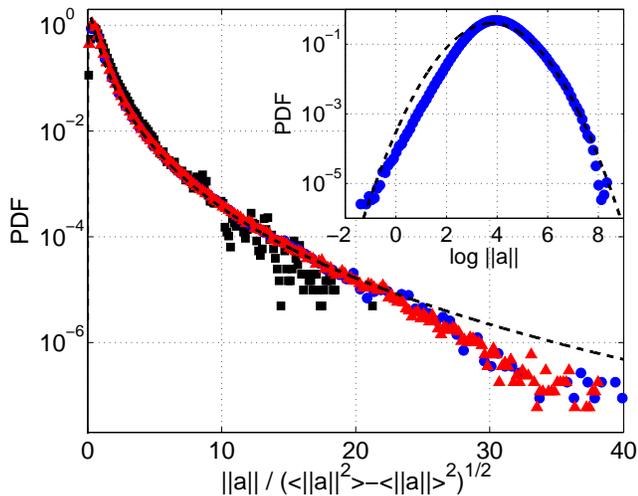}
\caption{PDF of the acceleration magnitude for $R_\lambda=285$ ({\small $\blacksquare$}), 485 ({\large $\blacktriangle$}), 690 ({\Large $\bullet$}). The dashed line is a log normal distribution of variance 1. Insert: PDF of $\log ||\mathbf a||$ ({\Large $\bullet$}: experiment at $R_\lambda=690$, dashed line: log-normal distribution of variance 1)}
\label{fig-amagdist}
\end{figure}

Our unique setup enables us to make the first experimental estimation of the acceleration magnitude at high Reynolds numbers. The insert in Fig.~\ref{fig-amagdist} presents the distribution of the logarithm of the magnitude at $R_\lambda=690$.  The experimental PDF of $\log\|\mathbf{a}\|$ resembles the one of Fig.~19 from~\cite{Yeung89} obtained from DNS at much lower Reynolds number ($R_\lambda=90$). The experimental curves are compared to a log-normal distribution,
\begin{equation}
P_{||a||}(a)=\frac{1}{as\sqrt{2\pi}}\exp\left(-\frac{(\ln (a/m))^2}{2s^2}\right)\, ,
\end{equation}
with $s=1$ (dashed line).
As can be seen in the inset of Fig.~\ref{fig-amagdist}, at large values of the acceleration, the curve is very close to the log-normal distribution while for low values it displays a wider tail than the log-normal, possibly exponential. However, the departure from log-normal at low values occurs for values of the acceleration which are much smaller than the average acceleration magnitude. Small values of acceleration are highly sensitive to experimental noise.
Indeed, a synthetic acceleration vector drawn from a genuine lognormal magnitude shows such a tail as soon as it is corrupted by noise. Thus this tail may not be representative of the real PDF.

Figure~\ref{fig-amagdist} displays the PDF of the acceleration magnitude for three Reynolds numbers. Despite the low statistics, at $R_\lambda=285$ it seems that the PDF decreases a little faster for large accelerations than in the two other data sets. For the two higher Reynolds numbers, the shape does not seem to evolve and the two curves are superimposed. All three curves compare well to a lognormal distribution of unit variance for normalized values lower than 25.

Now let us assume that the distribution of the acceleration magnitude is indeed log-normal, and that the acceleration vector is isotropic (an assumption which seems to be valid at the highest Reynolds numbers). Then one can compute the functional form of the acceleration components by
\begin{equation} 
P(a_i)=\frac{1}{2}\int_{|a_i|}^{\infty}\frac{P_{||a||}(a)}{a}\textrm{d}a
\end{equation}
where $P_{||a||}$ is the PDF of the magnitude~\cite{Pope}. If $P_{||a||}$ is log-normal, then the PDF of the components should be
\begin{equation}
P(a_i)=\frac{\exp(s^2/2)}{4m}\left[1-\textrm{erf}\left(\frac{\ln \left(\frac{|a_i|}{m}\right)+s^2}{\sqrt{2}s}\right)\right]\, .
\end{equation}
The parameter $m$ accounts only for the change in the component variance ($m=\sqrt{{3}/{e^{2s^2}}}$ for variance 1). The $s$ parameter determines the shape of the PDF. For the highest
 Reynolds numbers, the value $s=1$ gives very good agreement with the experimental PDFs of the acceleration components for values of $|a_i|$ up to 25 times the standard deviation (Fig.~\ref{fig-pdfacc}, solid line). 
For the highest values of the acceleration component, we observe a departure of the experimental curve from the log-normal behavior. The origin of this departure remains unclear.  It could be due to a real hydrodynamic effect, but may also be an artifact of the measurement process; for instance, the particles may not follow the very high acceleration events and thus underestimate the extreme tails. { The rather low statistics at the highest values of acceleration can also be responsible for the departure}

Under the same hypotheses, the PDF of the acceleration vector would be 
\begin{equation}
P(a_x,a_y,a_z)=\frac{P_{\|a\|}(\sqrt{a_x^2+a_y^2+a_z^2})}{4\pi (a_x^2+a_y^2+a_z^2)}
\end{equation}
so that  one can also compute the conditional average $\langle a_y^2 | a_z\rangle$. It is compared to the measurement at $R_\lambda=690$ in Fig.~\ref{fig-jPDFmoycond} (solid line). The agreement with the data is very good up to ten times the {\it rms} value of the acceleration component. For higher values, the data is lower than the log-normal result. Nevertheless, the log-normal curve remains close to the data, showing that the log-normal distribution of variance one seems to describe the data for moderate values of the acceleration.

\section{Time dynamics}

\begin{figure}
\includegraphics[width=8.5cm]{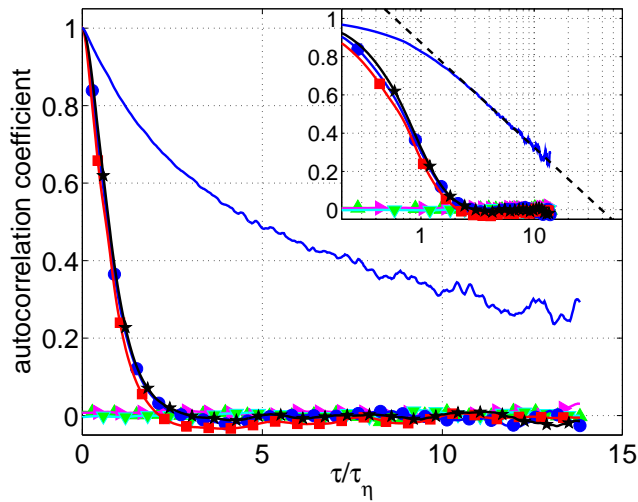}
\caption{Correlation curves at $R_\lambda=690$. {\large $\blacktriangle$} cross-correlations of the acceleration components. {\Large $\bullet$}, {\large $\star$}, {\small $\blacksquare$} are the autocorrelations of $a_x$, $a_y$ and $a_z$ respectively $\tau_\eta=\sqrt{\nu/\epsilon}$ is the Kolmogorov time ($\epsilon$ is the energy dissipation rate and $\nu$ is the kinematic viscosity). The line without symbols is the autocorrelation of the acceleration magnitude. Insert: semi-logarithmic plot; the dashed line is a extrapolation of the magnitude correlation.}
\label{fig-cor}
\end{figure}

The auto- and cross-correlations of the acceleration components are displayed in Fig.~\ref{fig-cor}. The cross-correlations of the different components are seen to be zero within the experimental uncertainty. The autocorrelations are very close to each other. Small differences are nevertheless observed, in particular between the horizontal ($x$, $y$) components and the vertical one ($z$). The auto-covariance of $a_z$ reaches zero at 2.2~$\tau_\eta$ (comparable to the values reported for DNS at low $R_\lambda$ in~\cite{Yeung89,Yeung97}) and the other two components at 3.0~$\tau_\eta$. We observed that the crossing times are almost independent of the Reynolds number. 

Figure~\ref{fig-cor} also displays the autocorrelation of the acceleration magnitude (solid line). The decrease of the covariance of the magnitude is seen to be much slower than the component correlation (the size of the measurement volume does not allow us to observe the full decrease of the magnitude covariance). This feature was observed in DNS at low Reynolds number by Yeung~\cite{Yeung89,Yeung97} and for velocity increments by Mordant {\it et al.}~\cite{Mordant02b}. 
At large $\tau$, a bias due to the finite measurement volume is expected to depress the autocorrelation~\cite{Crawford}. This bias will strongly affect the autocorrelation of the magnitude since it decays very slowly. Nevertheless, one can extrapolate the decrease in order to obtain an estimate of the characteristic time (Fig.~\ref{fig-cor}). We found a zero crossing time of $40\tau_\eta$, which is most likely underestimated because of the bias. This time is one seventh of the rotation period of the disks. Mordant {\it et al.} reported, for velocity increments, times that are close to one third of the rotation period~\cite{Mordant02b,Mordant04}. The dynamics of the acceleration magnitude was observed to have time scales comparable to those of the energy injection. A similar result was observed by Pope for low Reynolds number simulations~\cite{Pope90a}. This observation is quite striking as the components have been observed to evolve at the dissipation time scale as expected in the framework of the Kolmogorov 1941 theory. 


In summary, we have provided experimental evidence of the complex structure of the Lagrangian acceleration. The three components of the acceleration are not independent as in most stochastic models. We observed that the PDF of the acceleration magnitude is close to a log-normal distribution. Assuming isotropy, we showed that a log-normal distribution reproduces fairly accurately the experimental observations, including the dependence of the components. The acceleration also displays a rich dynamics with two different time scales of correlation: a short time for the direction and a longer one for the magnitude. Intense vortices are objects that fit well with the previous observations. They can account for the strong intermittency of the acceleration components. They induce simultaneous high accelerations in different components. If their lifetime is long enough, they can be responsible for the two time scales, since the magnitude of the acceleration remaining unchanged for times much longer than
the vortex rotation rate. 

\begin{acknowledgments} 
This work was supported by NSF grant \# 9988755.
\end{acknowledgments} 
\bibliographystyle{apsrev}
\bibliography{PRL.bib}

\end{document}